\documentclass[floatfix,twocolumn,reprint,superscriptaddress,aps,prl]{revtex4-1}
\usepackage{graphics}
\usepackage[dvips]{graphicx}
\usepackage{dcolumn}
\usepackage{bm}
\usepackage{subfigure}
\usepackage{color}
\usepackage{amsmath,amssymb,amsfonts}
\usepackage[linktocpage,colorlinks=true,linkcolor=blue,citecolor=blue]{hyperref}

\begin{document}


\title{Lattice operators from discrete hydrodynamics}

\author{Rashmi Ramadugu}
\author{Sumesh P.Thampi}
\affiliation{Engineering Mechanics Unit, Jawaharlal Nehru Centre for Advanced Scientific Research, Bangalore 560064, India}
\author{Ronojoy Adhikari} 
\affiliation{The Institute of Mathematical Sciences, CIT Campus, Chennai 600113, India}
\author{Sauro Succi} 
\affiliation{ Istituto Applicazioni Calcolo, CNR Roma - via dei Taurini 9, 00185, Roma, Italy, EU}
\author{Santosh Ansumali}
\affiliation{Engineering Mechanics Unit, Jawaharlal Nehru Centre for Advanced Scientific Research, Bangalore 560064, India}

\begin{abstract}

We present a general scheme to derive lattice differential operators from the discrete 
velocities and associated Maxwell-Boltzmann distributions used in lattice hydrodynamics. 
Such discretizations offer built-in isotropy and recursive techniques to increase the convergence order. 
This provides a simple and elegant procedure to derive isotropic and accurate discretizations of differential operators,
which are expected to apply across a broad range of problems in computational physics.
\end{abstract}

\maketitle
The development of efficient and accurate lattice versions of differential operators is a central theme of computational physics. 
Indeed, the numerical solution of any classical or quantum field theory requires the development of discrete differential operators in order to solve the partial differential equations associated with the continuum theory. Isotropy of the differential operators in continuum space is often lost in developing the corresponding discrete operators. This inability to perform discrete operations satisfying the inherent isotropy of continuum space may reflect severely on numerical simulations of physical problems. It is therefore desirable to develop discrete operators which retain as many symmetries as possible of their continuum counterparts. Here, we address this issue with specific regard to isotropy, and show that progress in lattice hydrodynamic simulations, naturally provides a strategy to develop such operators in full generality, beyond the original realm of hydrodynamics.

Finite difference schemes remain one of the most popular method of discretising differential operators. 
Though the accuracy of the scheme can be improved by increasing the stencil size, discrete operations are generally restrained 
to the principal directions of the lattice (coordinate directions on a rectangular grid), often neglecting the grid points along other directions. 
For example, in order to calculate the curl or Laplacian of a field, very often, only information available at principal directions is used. 
This leads to a loss of information which deteriorates the accuracy of the discrete operation, isotropy in the first place. 
Use of larger stencils with next-nearest neighbors may offer significant improvements without degrading efficiency to any significant extent. 
This issue has been addressed before in the literature. 
For instance, mimetic discretizations have been developed to recover the properties of underlying continuum theory \cite{Bochev2006, *Shashkov1999}. 
The isotropy of Laplacian operators has been the object of several previous studies \cite{Kumar2004, *Patra2005} and a specific illustration of the general method proposed was presented in \cite{ours}.
In this Letter, we show that the same procedure can be applied in full generality to a broad class
of differential operators, such as gradient, divergence and curl, which play a central role across virtually
all areas of computational physics.  

\begin{figure}
\begin{center}
\includegraphics[width=0.7\linewidth]{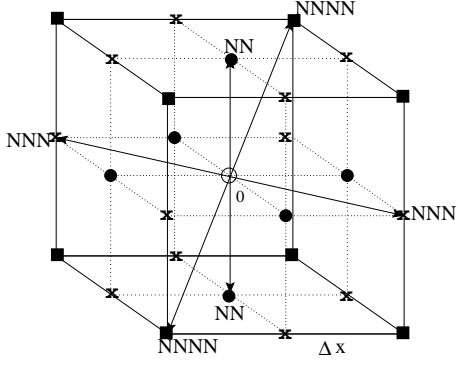}
\caption{\label{fig:lapcube} Points on a cubic cell of dimensions $2\Delta x \times 2\Delta x \times 2\Delta x$. $0$ is the point of interest and will be represented by $\mathbf{r}$. Here NN represent the nearest neighbors, NNN the next nearest neighbors, and NNNN the next next nearest neighbors, marked by $\circ, \bullet, \boldsymbol{\times}$ and {\tiny $\blacksquare$} respectively.}
\end{center}
\end{figure}

\textit{Discrete operators from lattice kinetic theory}:
Consider a unit cell of dimensions $2\Delta x \times 2\Delta x \times 2\Delta x$ as shown in Fig.\ (\ref{fig:lapcube}), generating a standard uniform grid in Cartesian coordinates. The center of the cube is the point of interest and will be denoted as $\mathbf{r}$. Neighboring points on the grid vary in distance and can be classified as  nearest neighbors - NN, next nearest neighbors - NNN, and next-next-nearest neighbors - NNNN, as highlighted in Fig.\ (\ref{fig:lapcube}). Let the vectors pointing to each of these points be denoted by $\mathbf{c}_i$, where $i$ represents any point on the grid. We include $\mathbf{c}_0$ which is a zero vector at $\mathbf{r}$. Let us also define weights, $w_i$, associated with these different points on the lattice. To be more precise, within the context of lattice kinetic theory, $\mathbf{c}_i$ represent a set of discrete speeds which move the information from the center point $\mathbf{r}$ to the $i$-th neighbor $\mathbf{r}+\mathbf{d}_i$ in a time-step $\Delta t$, according to the light-cone rule, $\mathbf{d}_i = \mathbf{c}_i \Delta t$. In the following, we shall take $\Delta t =1$, so that the discrete displacements $\mathbf{d}_i$ can be identified with the discrete speeds $\mathbf{c}_i$.

We begin with a D$n$Q$m$ lattice hydrodynamic model in $n$ dimensions with $m$ discrete velocities. It is well known from the lattice hydrodynamics literature that, in order to preserve isotropy of discrete space up to fourth order in the lattice tensors, it is necessary to have \cite{Hudong2008, *Mauro2007}
\begin{align}
\sum_{i} w_{i} &= 1 \label{eqn:ISO1} \\
\sum_{i} w_{i} c_{i,\alpha} c_{i,\beta} &= T \delta_{\alpha \beta} \label{eqn:ISO2} \\
\sum_{i} w_{i} c_{i,\alpha} c_{i,\beta} c_{i,\gamma} c_{i,\lambda} &= T^2 \Delta_{\alpha \beta \gamma \lambda}^{(4)}
\label{eqn:ISO3}
\end{align}
where Greek indices label Cartesian directions and $\Delta_{\alpha \beta \gamma \lambda}^{(4)} = \delta_{\alpha\beta} \delta _{\gamma\lambda} + \delta_{\alpha\lambda} \delta _{\gamma\beta} + \delta_{\alpha\gamma} \delta _{\beta\lambda}$. We choose $\Delta x=1$ such that $T = 1/3$, a lattice-dependent constant. It is well known that the lattice formulation of kinetic theory provides a computationally efficient method to solve conservation equations and is well established for the hydrodynamics \cite{SucciBook}. A discrete form of the Maxwell-Boltzmann velocity distribution \cite{Maxwell1860, *Boltzmann1872} is used in this formulation, which preserves the isotropy of space to fourth order.
Higher order stencils may be used to obtain 6th order accuracy \cite{kyu2010, *chikata2010}.

We now introduce the method of generating various discrete operators which preserves isotropy upto fourth order. Isotropy at higher orders needs larger stencils. Now consider the general operator ($\mathbf{s}\equiv\boldsymbol{\nabla}$)
\begin{equation}
 f_{\mathbf{s}} \equiv \frac{1}{T} \sum w_i e^{\mathbf{c}_i \cdot \mathbf{s}}
\end{equation}
that acts on a field $\psi({\bf r})$ discretely defined on a lattice as shown in Fig.\ (\ref{fig:lapcube}). As a result,
\begin{equation}
 f_{\mathbf{s}} \left[\psi({\bf r})\right] = \frac{1}{T} \sum w_i e^{\mathbf{c}_i \cdot \mathbf{s}} \psi(\mathbf{r}) = \frac{1}{T} \sum w_i \psi(\mathbf{r}+\mathbf{c}_i).
\end{equation}
We illustrate below how $f_{\mathbf{s}}$ can act as a generating function to construct several discrete operators.

Consider the following two transformations $\mathcal{F}(\psi) = 2\left(f_{\mathbf{s}} - f_{\mathbf{0}}\right)\psi(\mathbf{r})$ and $\boldsymbol{\mathcal{G}}(\psi) = \frac{df}{d\mathbf{s}}\psi(\mathbf{r})$
%
%
or equivalently
\begin{align}
\mathcal{F}(\psi) &= \frac{2}{T} \sum_{i=1}^{N} w_{i} \left[\psi(\mathbf{r}+\mathbf{c}_{i}) - \psi(\mathbf{r})\right]
\label{eqn:laptrans} \\
\boldsymbol{\mathcal{G}}(\psi) &= \frac{1}{T} \sum_{i=1}^{N} w_{i} \mathbf{c}_{i} {\psi}(\mathbf{r}+\mathbf{c}_{i})
\label{eqn:gradtrans}
\end{align}
where $N$ is the total number of neighboring points considered. Taylor expanding these expressions and using the symmetries of Eqs.\ (\ref{eqn:ISO1} - \ref{eqn:ISO3}), we obtain
\begin{align}
\mathcal{F}(\psi) = \left(1 + \frac{T}{4} \nabla^2 \right) \nabla^2 \psi ~ \textnormal{and}~
\boldsymbol{\mathcal{G}}(\psi) = \left(1 + \frac{T}{2} \nabla^2\right)  \boldsymbol{\nabla} \psi. \nonumber
\end{align}
%
%
%
%
These expressions
may be solved for $\nabla^2 \phi$ and $\boldsymbol{\nabla} \phi$ respectively by inverting the linear operators.
%
%
We retain only the leading order terms and show how these expressions may be used to obtain isotropic discrete operators and how a recursive technique can be employed to obtain higher order accurate discrete 
operators.

Considering only the leading order terms of the inverted linear operators
, we find that
\begin{align}
 \nabla^2 \psi &= \frac{2}{T} \left[\sum_{i=1}^{N} w_{i} \psi(\mathbf{r}+\mathbf{c}_{i}) - \psi(\mathbf{r})\right] + O(\nabla^4),
\label{eqn:lap1}\\
 \boldsymbol{\nabla} \psi &=  \frac{1}{T} \sum_{i=1}^{N} w_{i} \mathbf{c}_{i} \psi(\mathbf{r}+\mathbf{c}_{i}) + O(\nabla^3).
\label{eqn:grad1}
\end{align}
These expressions preserve isotropy at the leading order. The former expression has been reported earlier in \cite{ours}. The latter is also suggested and used in various contexts \cite{Rotenberg2008, *sumeshlb}. Here we explain the underlying connection to lattice hydrodynamics and present the corresponding expressions for the divergence and the curl. Hence, for a vector field, $\boldsymbol{\phi}({\bf r})$,
\begin{align}
\boldsymbol{\nabla} \cdot \boldsymbol{\phi} &=  \frac{1}{T} \sum_{i=1}^{N} w_{i} \mathbf{c}_{i} \cdot \boldsymbol{\phi}(\mathbf{r}+\mathbf{c}_{i}) + O(\nabla^3)
\label{eqn:div1}\\
\boldsymbol{\nabla} \wedge \boldsymbol{\phi} &=  \frac{1}{T} \sum_{i=1}^{N} w_{i} \mathbf{c}_{i} \wedge \boldsymbol{\phi}(\mathbf{r}+\mathbf{c}_{i}) + O(\nabla^3) .
\label{eqn:curl1}
\end{align}
These expressions are remarkable because they display isotropy up to leading order error, with an error coefficient of order $O(T)$. Any lattice with suitable weights which satisfies the conditions in Eq.\ (\ref{eqn:ISO1} - \ref{eqn:ISO3}) will provide an expression for the discrete operators and will ensure isotropy. For example the weights used in the lattice hydrodynamics may be used. These weights are lattice analogues of the  Maxwell-Boltzmann equilibrium in continuum velocity space. 

A recursive algorithm can be developed now to remarkably improve the accuracy of these operators. Considering only the next order terms of the inverted linear operators
, we have
\begin{align}
\nabla^2 \psi &= \mathcal{F}(\psi) - \frac{T}{4} \mathcal{F} (\mathcal{F}(\psi))
\label{eqn:lap2} \\
 \boldsymbol{\nabla} \psi &= \boldsymbol{\mathcal{G}}(\psi) - \frac{T}{2} \mathcal{F} (\boldsymbol{\mathcal{G}}(\psi)) .
\label{eqn:grad2} 
\end{align}
In the explicit form, these expressions are
\begin{align}
 \nabla^2 \psi &= \frac{2}{T} \left[\sum_{i=1}^{N} w_{i} \psi(\mathbf{r}+\mathbf{c}_{i}) - \psi(\mathbf{r})\right] \nonumber\\ &- \frac{1}{2} \left[\sum_{i=1}^{N} w_{i} \mathcal{F}(\psi(\mathbf{r}+\mathbf{c}_{i})) - \mathcal{F}(\psi(\mathbf{r}))\right]
\label{eqn:laplead} \\
 \boldsymbol{\nabla} \psi &=  \frac{1}{T} \sum_{i=1}^{N} w_{i} \mathbf{c}_{i} \psi(\mathbf{r}+\mathbf{c}_{i}) \nonumber\\ &-  \left[\sum_{i=1}^{N} w_{i} \boldsymbol{\mathcal{G}}(\psi(\mathbf{r}+\mathbf{c}_{i})) - \boldsymbol{\mathcal{G}}(\psi(\mathbf{r}))\right] .
\label{eqn:gradlead}
\end{align}
%
%
By employing this approximation for the next order term, we are essentially following a predictor-corrector method, however with a much simpler implementation.
The transformation $\boldsymbol{\mathcal{G}}(\psi)$ can be easily manipulated to obtain the divergence and curl operators. For a vector function $\boldsymbol{\phi}({\bf r})$ defined on a grid, the generating function can be manipulated as $\displaystyle \mathcal{D}(\boldsymbol{\phi}) = \frac{df}{d\mathbf{s}}\cdot \boldsymbol{\phi}({\mathbf{r}})$ and $\displaystyle \mathcal{C}(\boldsymbol{\phi}) = \frac{df}{d\mathbf{s}}\wedge \boldsymbol{\phi}({\mathbf{r}})$ to obtain
\begin{align}
\mathcal{D} (\boldsymbol{\phi}) &=  \frac{1}{T} \sum_{i=1}^{N} w_{i} \mathbf{c}_{i} \cdot \boldsymbol{\phi}(\mathbf{r}+\mathbf{c}_{i})
\\
\mathcal{C} (\boldsymbol{\phi}) &=  \frac{1}{T} \sum_{i=1}^{N} w_{i} \mathbf{c}_{i} \wedge \boldsymbol{\phi}(\mathbf{r}+\mathbf{c}_{i}) .
\end{align}
which provide the expressions for divergence and curl:
\begin{align}
\boldsymbol{\nabla} \cdot \boldsymbol{\phi} &=  \mathcal{D}(\psi) - \frac{T}{2} \mathcal{F} (\mathcal{D}(\psi))
\label{eqn:divlead}\\
\boldsymbol{\nabla} \wedge \boldsymbol{\phi} &=  \mathcal{C}(\psi) - \frac{T}{2} \mathcal{F} (\mathcal{C}(\psi)) .
\label{eqn:curllead}
\end{align}
By employing the recursive technique we obtain a higher order accurate method but the leading order error, which is $O(\nabla^5)$ in case of Eq.\ (\ref{eqn:gradlead}), (\ref{eqn:divlead}) and (\ref{eqn:curllead}), is not isotropic. As mentioned earlier, it is not possible to get the isotropic error beyond fourth order with this lattice. However, it is possible to go to larger stencils and obtain isotropy at the desired level in the same formulation. Double-differential operators, such as $\nabla(\nabla \cdot) $, $\nabla \cdot (\nabla \wedge) $, $\nabla \wedge (\nabla \wedge) $ and similar, may be derived likewise from the generating function using $d^2f/d\mathbf{s}^2$.

In short, for any transformation $\mathcal{K}(\boldsymbol{\phi})$ which at leading order provides a differential operation on $\boldsymbol{\phi}$, say $K[\boldsymbol{\phi}]$, it is possible to write
\begin{equation}
K[\boldsymbol{\phi}] \approx \mathcal{K}(\boldsymbol{\phi}) - \lambda \mathcal{F}\left[\mathcal{K}(\boldsymbol{\phi})\right]
\end{equation}
providing a method to perform the same differential operation. In the above expression, $\lambda = T/4$ for the Laplacian, and $\lambda = T/2$ for gradient, divergence and curl operators. Discrete operators can now be derived using any standard D$n$Q$m$ models. Popular models and the associated weights in lattice hydrodynamics literature are listed in Table (\ref{tab:DnQm}).

\begin{table}
\begin{tabular}{|c|cc||c|c|c|c|}
\hline
 & $N$, &(for 2D) & D2Q9 &D3Q15 & D3Q19 & D3Q27 \\
\hline
 0 & 1 &(1) & 4/9 &2/9 & 1/3 & 8/27 \\
\hline
  NN & 6 &(4) &  1/9 &1/9 & 1/18 & 2/27 \\
\hline
  NNN & 12 &(4) & 1/36 &0 & 1/36 & 1/54 \\
\hline
   NNNN & 8 &(0) & 0 &1/72 & 0 & 1/216 \\
\hline
\end{tabular}
\caption{Popular models and the associated weight factors for various D$n$Q$m$ lattice hydrodynamics models. 
Values of $N$ for the two-dimensional D2Q9 model are given in brackets.}
\label{tab:DnQm}
\end{table}
%
%
%
\begin{figure*}
\subfigure[~Continuous]{\includegraphics[trim=0 50 0 40, clip, scale=0.37]{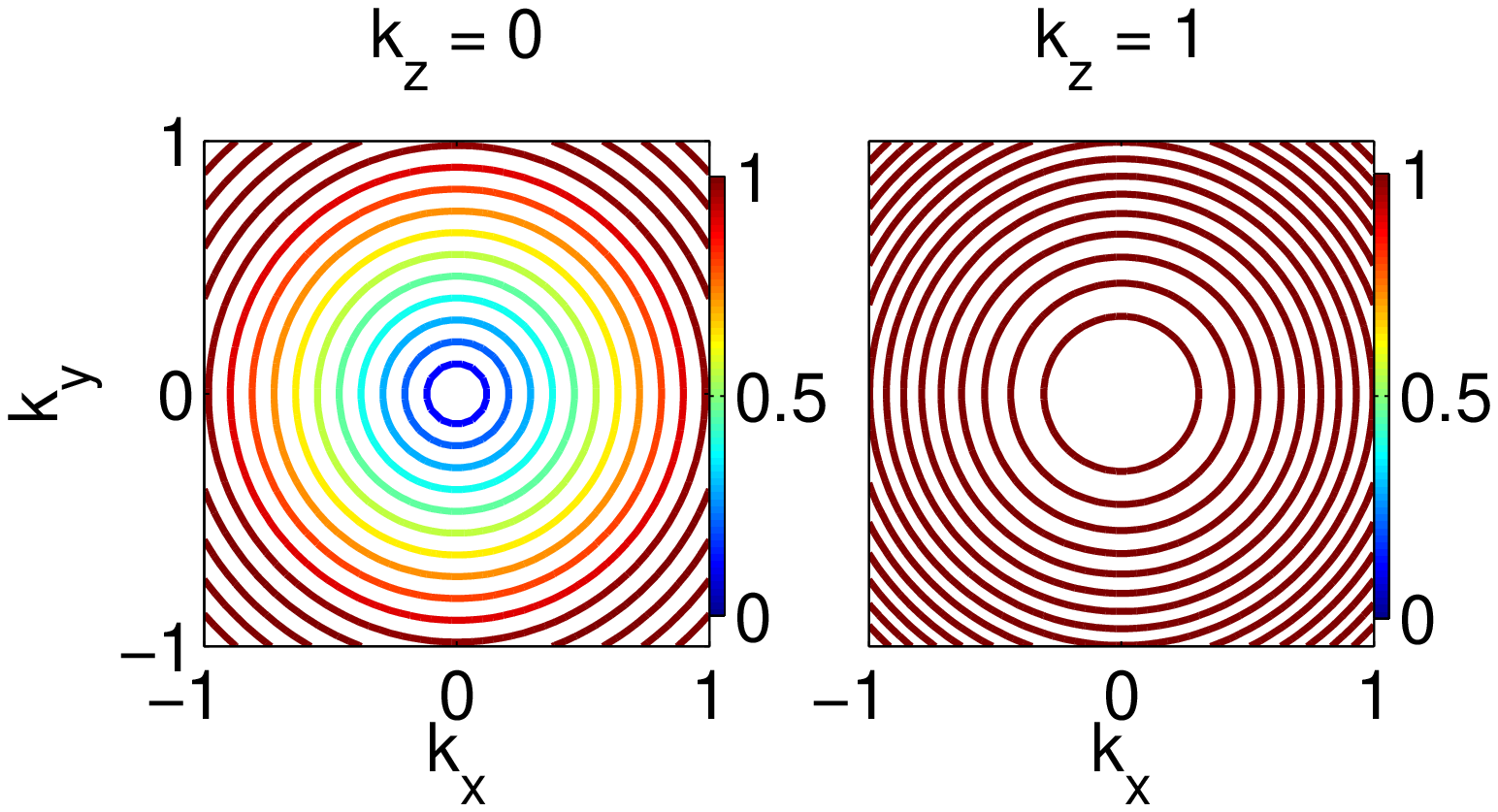}}
\subfigure[~CD]{\includegraphics[trim=0 50 0 40, clip, scale=0.37]{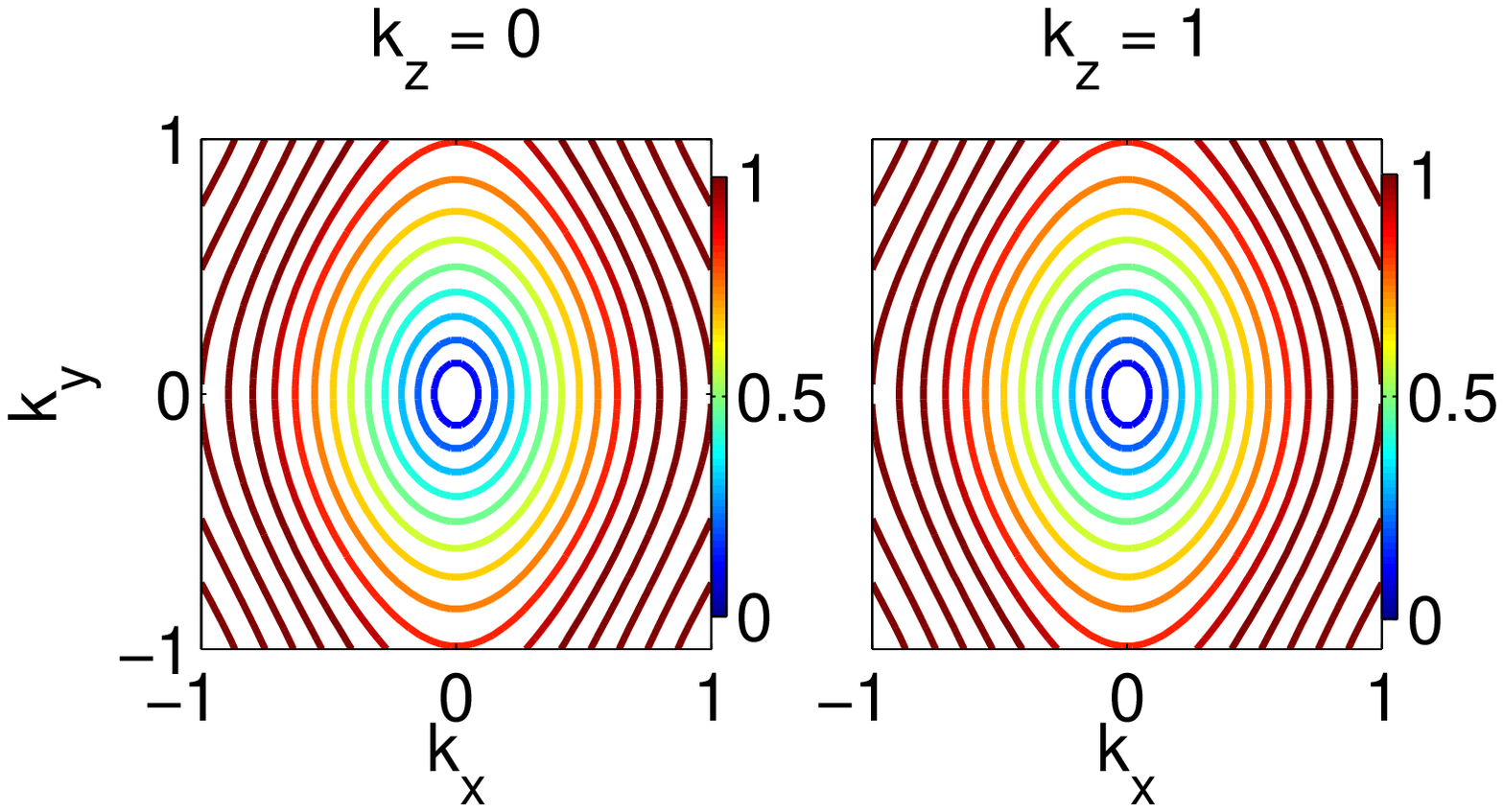}}
\subfigure[~D3Q15]{\includegraphics[trim=0 50 0 40, clip, scale=0.37]{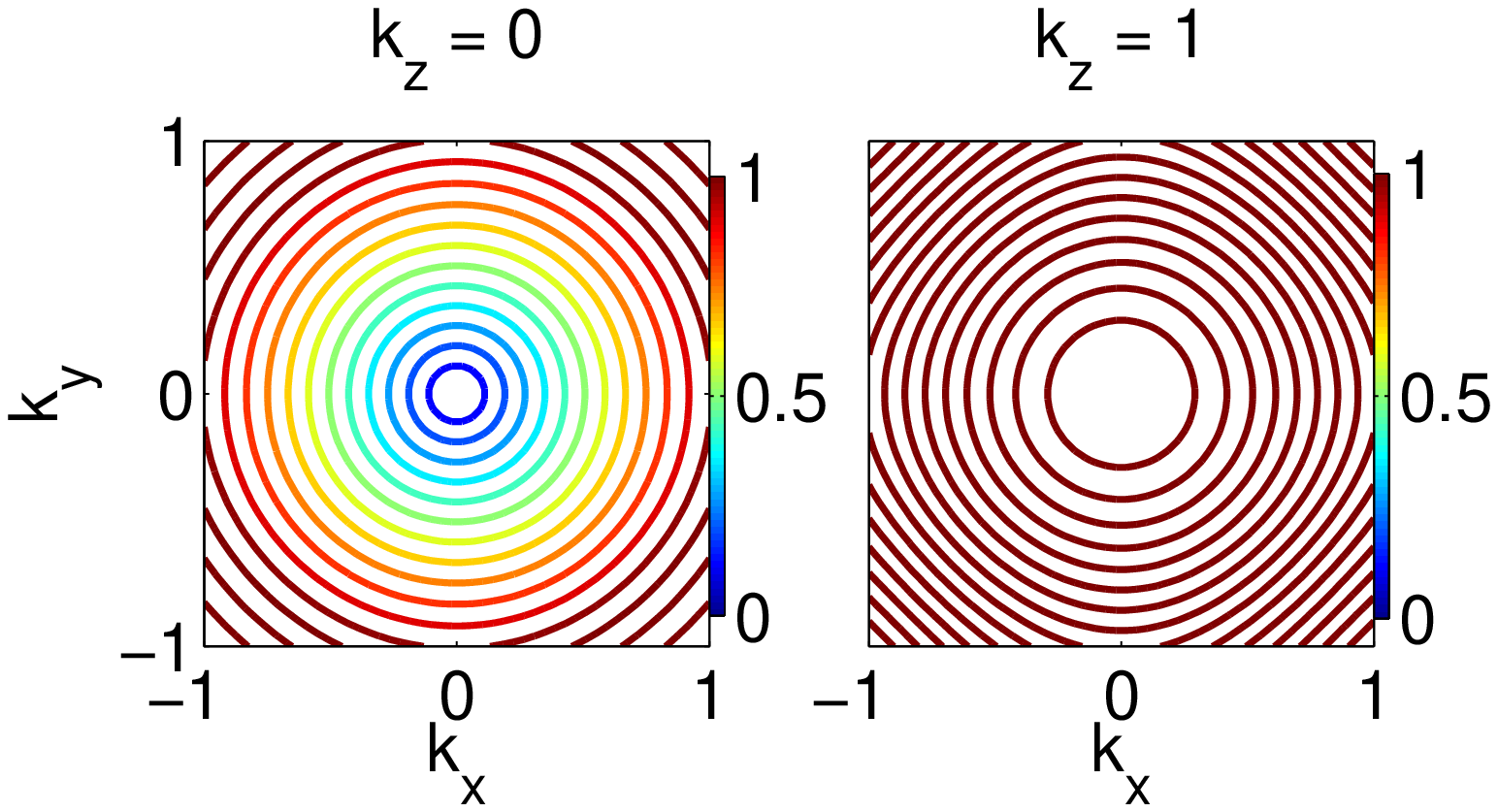}}\\
\vspace{-4mm}
\subfigure[~D3Q19]{\includegraphics[trim=0 50 0 40, clip, scale=0.37]{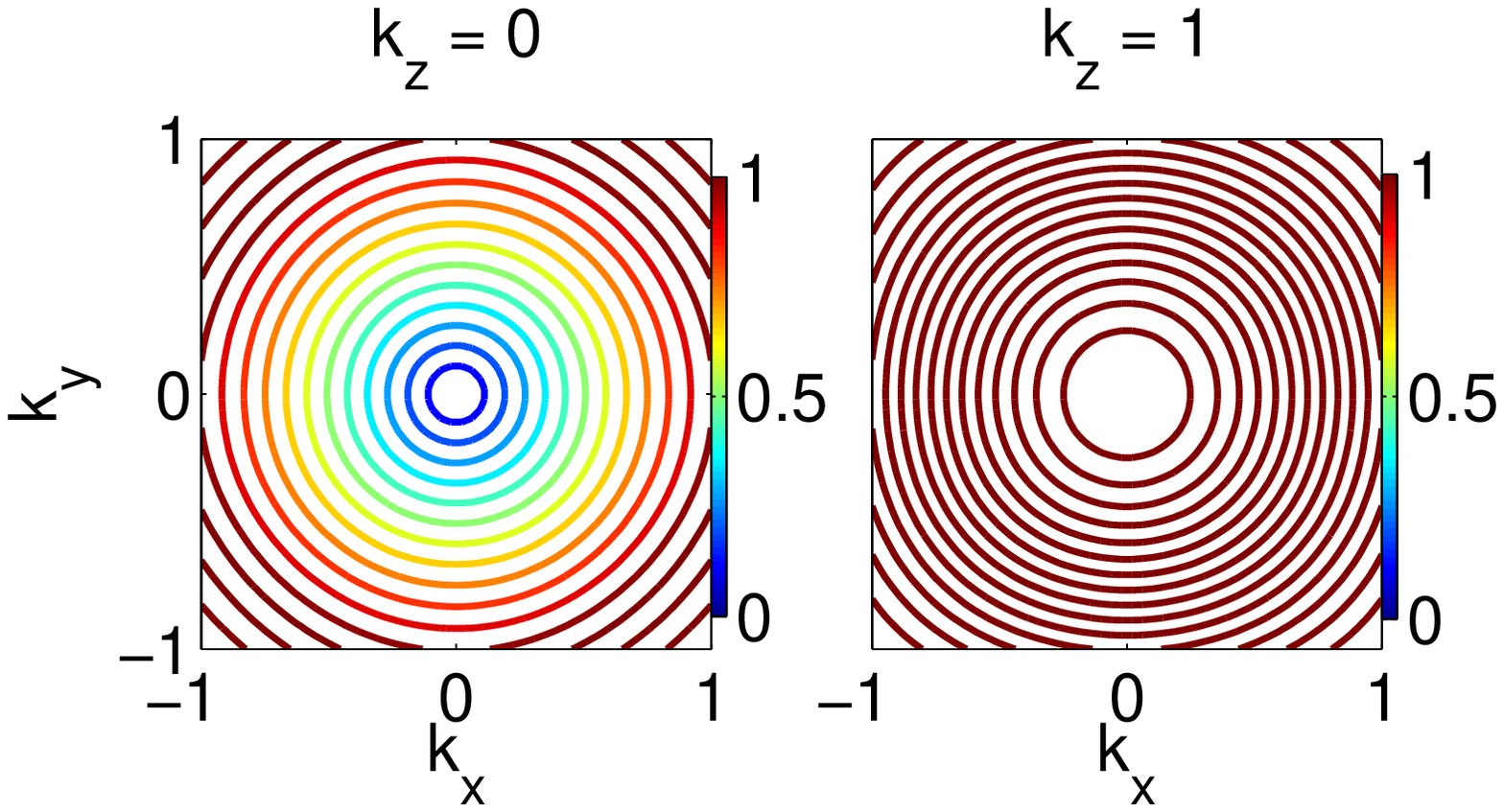}}
\subfigure[~D3Q27]{\includegraphics[trim=0 50 0 40, clip, scale=0.37]{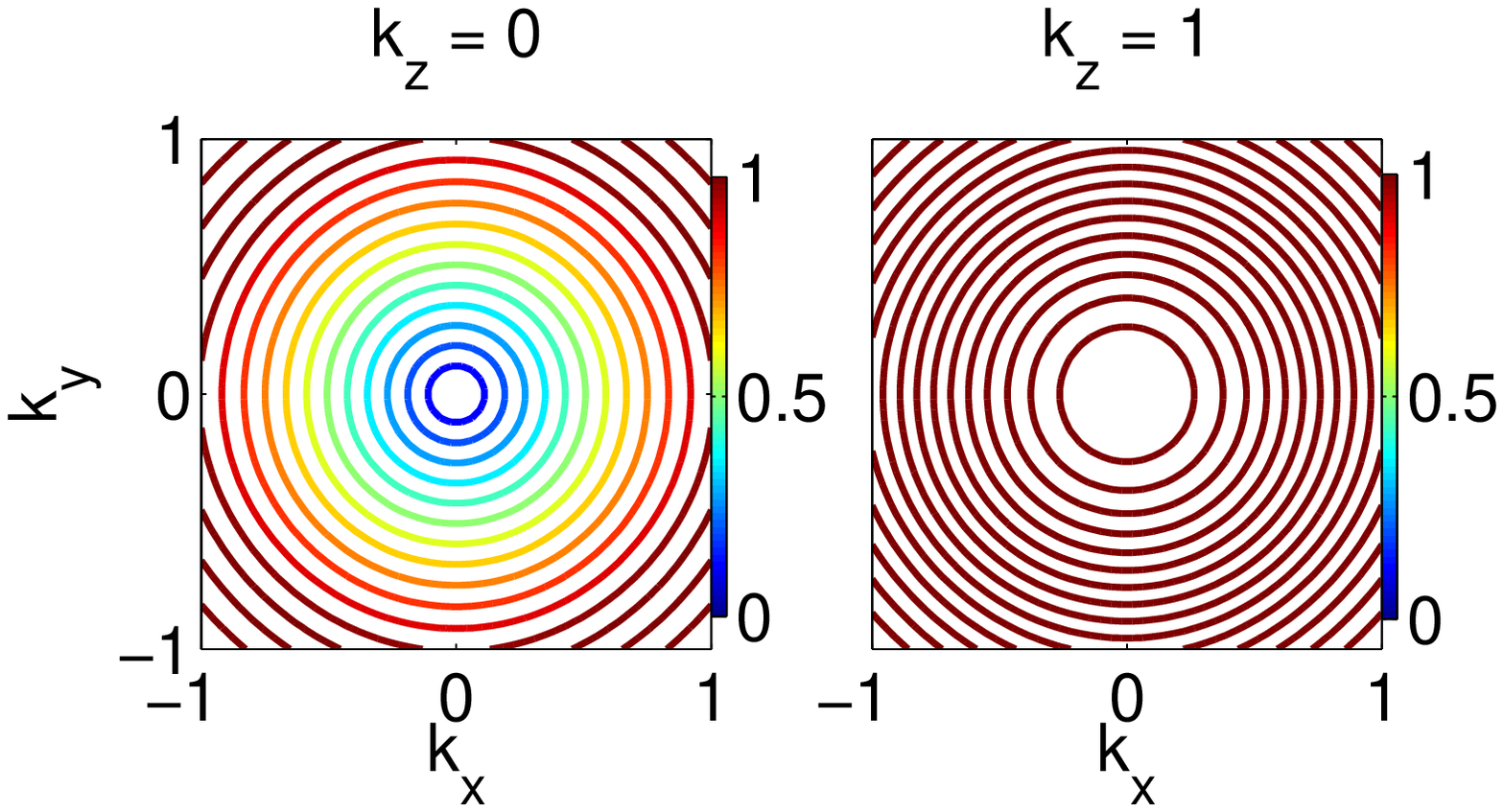}}
\caption{Contour plots in Fourier space of the discrete gradient operator $D(\mathbf{k})$, are shown for the planes $k_z=0$ and $k_z=1$ for (a) continuous (b) central difference, (c) D3Q15, (d) D3Q19 and (e) D3Q27 models. The present schemes invariably exhibit better isotropy. 
For clarity, the y-axis is omitted in $k_z = 1$ plots.}
\label{fig:contopr}
\end{figure*}

\textit{Results and Discussion}:
We next compare the accuracy and isotropy properties of the discrete operators, Eqns.\ (\ref{eqn:laplead} - \ref{eqn:gradlead}, \ref{eqn:divlead} - \ref{eqn:curllead}) that were derived in the last section. The discrete Fourier transform of the gradient operator is given by $D(\mathbf{k}) =  \sum_r \exp(-i\mathbf{k}. \mathbf{r})\boldsymbol{\mathcal{G}}(\psi)/\sum_r \exp(-i\mathbf{k}. \mathbf{r}) \psi(\mathbf{r})$. In the small wavelength limit, the corresponding expressions for different models and the standard second order central difference ($CD$) scheme may be written as follows. For clarity, only one component is shown below.
\begin{align}
 &D(\mathbf{k})_{\alpha}^{D2Q9} = ik_{\alpha}\left[1 - \frac{k^4}{36}-\frac{k_{\alpha}^4}{180} + O(k^6)\right]\nonumber
 \\
 &D(\mathbf{k})_{\alpha}^{D3Q15} = ik_{\alpha}\left[1-\left(\frac{k^4}{36}+\frac{k_{\alpha}^4}{180}+\frac{k_{\beta}^2 k_{\gamma}^2}{18}\right)+ O(k^6)\right]\nonumber
\\
 &D(\mathbf{k})_{\alpha}^{D3Q19} =  ik_{\alpha}\left[1 - \left(\frac{k^4}{36} + \frac{k_{\alpha}^4}{180} +\frac{k_{\beta}^2 k_{\gamma}^2}{36}\right) + O(k^6)\right]\nonumber
\\
 &D(\mathbf{k})_{\alpha}^{D3Q27} =  ik_{\alpha}\left[1 -\left(\frac{k^4}{36} + \frac{k_{\alpha}^4}{180}\right)+O(k^6)\right]\nonumber
\\
&D(\mathbf{k})_{\alpha}^{CD} = ik_{\alpha}\left[1 -\frac{k_{\alpha}^2}{6}+O(k^4)\right]\nonumber
\end{align}
where $k=|\mathbf{k}|$ and $(\alpha \neq \beta \neq \gamma)$ are the cartesian indices. For curl and divergence, the form 
of these operators remains the same. 
Note that repeated indices are not summed upon.

Contour plots of these operators in Fourier space are plotted in Fig.\ (\ref{fig:contopr}). 
The error involved in these calculations may be estimated by comparing with the analytical value $|\mathbf{k}|$.
Our schemes always provide better isotropy compared to the standard finite difference method at small wave numbers.
Such a comparison for Laplacian operators can be found in \cite{ours}. We now apply these derived discrete differential operators to various test functions and compare them with standard difference methods in the literature.

\begin{figure*}
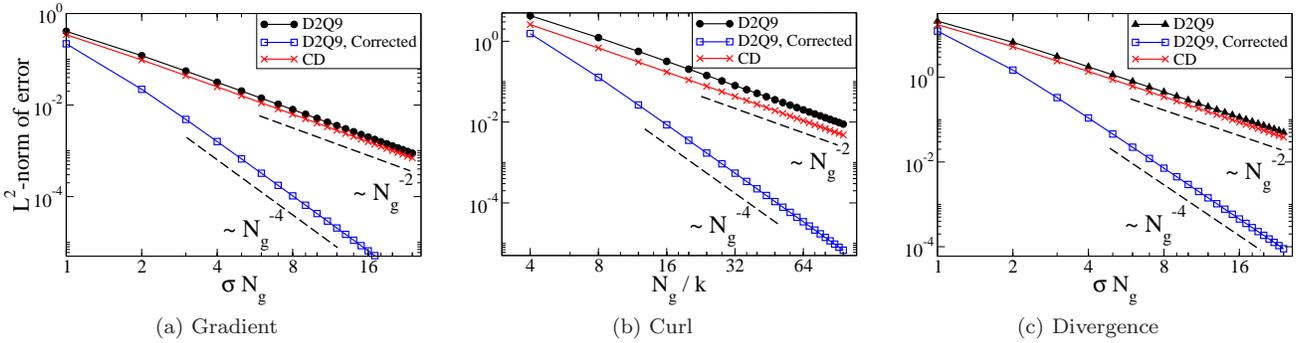

\centering
\subfigure[~Gradient]{\includegraphics[trim=0 0 0 0, clip, scale =0.23]{gradl2.eps}
\label{fig:gradl2}}
\subfigure[~Curl]{\includegraphics[trim=0 0 0 0, clip, scale =0.23]{curll2.eps}
\label{fig:curll2}}
\subfigure[~Divergence]{\includegraphics[trim=0 0 0 0, clip, scale =0.23]{divl2.eps}
\label{fig:divl2}}
\caption{$L^2$ -norms of the error for (a) discrete gradient operator applied on a Gaussian function, Eq. (\ref{eqn:gscalar}), (b) curl operator on Taylor-Green velocity field, Eq. (\ref{eqn:tgflow}), and (c) divergence operator on a Gaussian vector field,  in two dimensions are plotted as a function of grid spacing in log-log scale. Since the domain is of fixed size of unity for (a) and (c), abscissa is suitably scaled as $\sigma / \Delta x \equiv \sigma N_g$. We have used $\sigma=0.05$. In case of (b), $k_x = 3, k_y = 4$ and domain length $= 2 \pi$ are chosen, hence, abscissa is nondimensionalised as $(2 \pi/k)/\Delta x$. 
For clarity, the ordinate is labelled only in the left most panel. 
Dashed lines of slope 2 and 4 are plotted for eye-guiding purposes. 
The significant improvement of the recursive scheme may be noted in all cases.}
\label{fig:testfns}
\end{figure*}

We consider a two dimensional Gaussian as a test function,
\begin{equation}
\psi(x,y) = \exp\left[\frac{-(x^2 +y^2)}{2 \sigma^2}\right],
\label{eqn:gscalar}
\end{equation}
where $\sigma$ is the variance. 
%
%
In a square domain of length unity spanned by $N_g$ grid points in each direction, we compute the gradient using Eq.\ (\ref{eqn:grad1}) for a D2Q9 model. The gradient is also calculated using the modified scheme, Eq.\ (\ref{eqn:gradlead}), and the standard second order central difference formula. Comparing with the analytical expression for the gradient, $\boldsymbol{\nabla}\psi(x,y)^{\mathrm{analyt}}$,  $L^2-\textnormal{norm}$ of the error is defined as
\begin{align}
L^2-\textnormal{norm} = \frac{\sqrt{\sum_i^{N} \left[\boldsymbol{\nabla}({\psi})^{D2Q9} - \boldsymbol{\nabla}(\psi)^{\mathrm{analyt}}\right]^2}}{N_g}.
\end{align}
The results are illustrated in Fig.\ \ref{fig:gradl2}, where the error is plotted as a function of grid spacing, suitably nondimensionalised.
The increase in accuracy with the increase in number of grid points is apparent. The second order convergence of Eq.\ (\ref{eqn:grad1}) and the fourth order convergence of Eq.\ (\ref{eqn:gradlead}) may also be seen. While our lower order scheme maintains the isotropy property 
compared to the standard central difference method, the recursive scheme improves the accuracy considerably, i.e to fourth order.

Next we address the curl operator, whose discretization  is central  not only to hydrodynamics
but also to computational electromagnetics \cite{Chew1994, *Hanasoge2011}.
As a test  function, $\mathbf{u}(x,y)$, we choose,
\begin{equation}
\left.
\begin{aligned}
u_x &= -k_y \cos(k_x x) \sin(k_y y)\\
u_y &= k_x \cos(k_y y) \sin(k_x x)
\end{aligned}
\right\}
\label{eqn:tgflow}
\end{equation}
which represents the velocity field in a Taylor-Green flow. The vorticity, which is the curl of velocity field, is given by $k^2 \cos(k_x x)\cos(k_y y) \hat{\mathbf{z}}$ where $k^2 = k_x^2 + k_y^2$. The curl of this field is also obtained for a D2Q9 lattice model, using both lower order scheme and the improved scheme, Eqs.\ (\ref{eqn:curl1}) and (\ref{eqn:curllead}) and compared with the analytical expression. The error, defined using $L^2-\textnormal{norm}$ as earlier, is plotted in Fig.\ \ref{fig:curll2}. The curl, as calculated using the standard central difference scheme, is  also plotted in the same figure. It may be noted that, like for the gradient, the discrete expressions derived here provide significantly higher accuracy. Again, it may be noted that isotropy is in-built in the lower order operator.

We now compute the divergence of a Gaussian vector field, $\boldsymbol{\nabla}\psi(x,y)^{\mathrm{analyt}}$
%
using the D2Q9 model. Equations (\ref{eqn:div1}) and (\ref{eqn:divlead}) were used to compute the divergence of this function and compared with the analytical expression $\displaystyle \exp \left[-(x^2 + y^2)/2 \sigma^2 \right] \left[\frac{x^2 +y^2}{\sigma^4} - \frac{2}{\sigma^2}\right]$. The corresponding error, as $L^2$ norm, is plotted in Fig.\ \ref{fig:divl2}. The improvement in accuracy, as compared to a standard central difference operator, is also illustrated.

It may be mentioned that the Laplacian of a Gaussian function is calculated and the error, as $L_2$ norm, for the discrete operators, Eqns.\ (\ref{eqn:lap1}) and (\ref{eqn:laplead}), behaved in a similar fashion, witnessing the generality of the method.

Summarizing, we have shown that the stencils associated with the discrete-velocity schemes developed in lattice hydrodynamics, naturally 
provide an elegant, efficient and accurate procedure to formulate discrete isotropic versions of the most fundamental differential 
operators, such as gradient, curl, divergence and Laplacian. 
Furthermore, we have also shown that the accuracy of these operators can be systematically improved by means of a recursive iteration procedure. 
Application of these discrete operators to various smooth test functions, was shown to result in significantly improved accuracy, as compared to standard finite-difference operators. Finally, we wish to emphasize that, owing to its generality, the present method is expected
to apply to further classes of differential operators, such as the Dirac propagator and Wilson plaquettes
in lattice gauge theories \cite{creutz1985quarks}.

\bibliography{reference}

\end{document}